\documentclass[conference, a4paper]{IEEEtran}
\IEEEoverridecommandlockouts
\usepackage{cite}
\usepackage{amsmath,amssymb,amsfonts}
\usepackage{algorithmic}
\usepackage{graphicx}
\usepackage{textcomp}
\usepackage{xcolor}
\def\BibTeX{{\rm B\kern-.05em{\sc i\kern-.025em b}\kern-.08em
    T\kern-.1667em\lower.7ex\hbox{E}\kern-.125emX}}
\begin{document}

\title{Enhanced Security with Encrypted Vision Transformer in Federated Learning
\thanks{Identify applicable funding agency here. If none, delete this.}
}

\author{\IEEEauthorblockN{1\textsuperscript{st} Aso Rei}
\IEEEauthorblockA{\textit{Tokyo Metropolitan Univercity} \\
Tokyo, Japan \\
aso-rei@ed.tmu.ac.jp}
\and
\IEEEauthorblockN{2\textsuperscript{nd} Shiota Sayaka}
\IEEEauthorblockA{\textit{Tokyo Metropolitan Univercity} \\
Tokyo, Japan \\
shiota@tmu.ac.jp}
\and
\IEEEauthorblockN{3\textsuperscript{rd} Kiya Hitoshi}
\IEEEauthorblockA{\textit{Tokyo Metropolitan Univercity} \\
Tokyo, Japan \\
kiya@tmu.ac.jp}

}

\maketitle

\begin{abstract}
Federated learning is a learning method for training models over multiple participants without directly sharing their raw data, and it has been expected to be a privacy protection method for training data. In contrast, attack methods have been studied to restore learning data from model information shared with clients, so enhanced security against attacks has become an urgent problem. Accordingly, in this article, we propose a novel framework of federated learning on the bases of the embedded structure of the vision transformer by using the model information encrypted with a random sequence. In image classification experiments, we verify the effectiveness of the proposed method on the CIFAR-10 dataset in terms of classification accuracy and robustness against attacks.
\end{abstract}


\section{Introduction}
\noindent
Deep neural networks (DNNs) have been deployed in various applications. Training high-performance DNN models requires a huge amount of training data, and training data include sensitive information such as personal information in general. Accordingly, it is difficult to prepare an amount of data to train DNN models, so privacy-preserving methods for deep learning have become an urgent problem \cite{SIP-2021-0048, 8486525}.
Federated learning (FL) has been excepted as one of the solutions \cite{pmlr-v54-mcmahan17a}. FL is a type of distributed machine learning. It is a model learning method that reflects all clients’ data by sharing only the updated information of each local model without directly sharing each client’s training data. However, it has been pointed out to be vulnerable to state-of-the-attacks \cite{NEURIPS2019_60a6c400, 10.5555/3495724.3497145, Lu_2022_CVPR}. In particular, vision transformer (ViT) models \cite{ViT}, which are known to have a high performance, are highly vulnerable as discussed in \cite{Lu_2022_CVPR}. 

Therefore, various privacy-preserving methods have been proposed to enhance security in FL so far. Differential privacy\cite{TCS-042} is one of the state-of-the-art, in which the values of model parameters are hidden by adding noise with a specific distribution. However, there is a trade-off relation between the level of privacy protection and model performance, so if we want to strongly protect model parameters, the use of differential privacy degrades the performance of models.

\indent Accordingly, in this paper, we propose a novel framework for enhancing the security of ViT models in FL. In the proposed framework, focusing on the embedding structure of ViT, each updated local model is encrypted by using a random matrix generated with a secret key, which has been inspired by privacy-preserving deep learning with encrypted images for ViT\cite{HitoshiKIYA20232022MUI0001, 9760030, 9909972, jimaging8090233}. Encrypted local-model information is extracted from each encrypted model, and the encrypted local-model information is shared with clients to perform model integration directly in the encrypted domain. In experiments, the proposed method is demonstrated not only to maintain the same accuracy as that of FL without any encryption but to also enhance robustness against an attack called Attention Privacy Leakage (APRIL)\cite{Lu_2022_CVPR}, which aims to restore the visual information of training images from updated local-model information.

\section{Related Work}
\subsection{Federated Learning}
Before discussing the proposed method, we summarize the general procedure of FL.
\begin{quote}
 \begin{itemize}
  \item[i)] A server provider distributes an initial global model to all clients.
  \item[ii)] Each client updates the global model with their training data.
  \item[iii)] Each client sends the updated model to the server.
  \item[iv)] The server provider integrates the model information received from all clients and updates the global model.
  \item[v)] The server provider sends the updated global model to all clients.
  \item[vi)] Repeat ii) to v) multiple times.
 \end{itemize}
\end{quote}
FedAvg (Federated Averaging)\cite{pmlr-v54-mcmahan17a} and FedSGD (Federated Stochastic Gradient Descent)\cite{pmlr-v54-mcmahan17a} are typical methods for updating global models.
In FedAvg, each client computes the gradient of an image and updates the local model with the gradient. After that, the client sends the parameters of the local model to the server. The server integrates the model parameters from clients to update the global model. In FedSGD, the client computes gradients from an image and sends them to the server. The server updates the model based on the gradients from each client. The proposed method can be adapted to both methods.
\subsection{Vision Transformer}
\begin{figure}[bth]
    \centering
    \includegraphics[bb=0 0 806 500,scale=0.25]{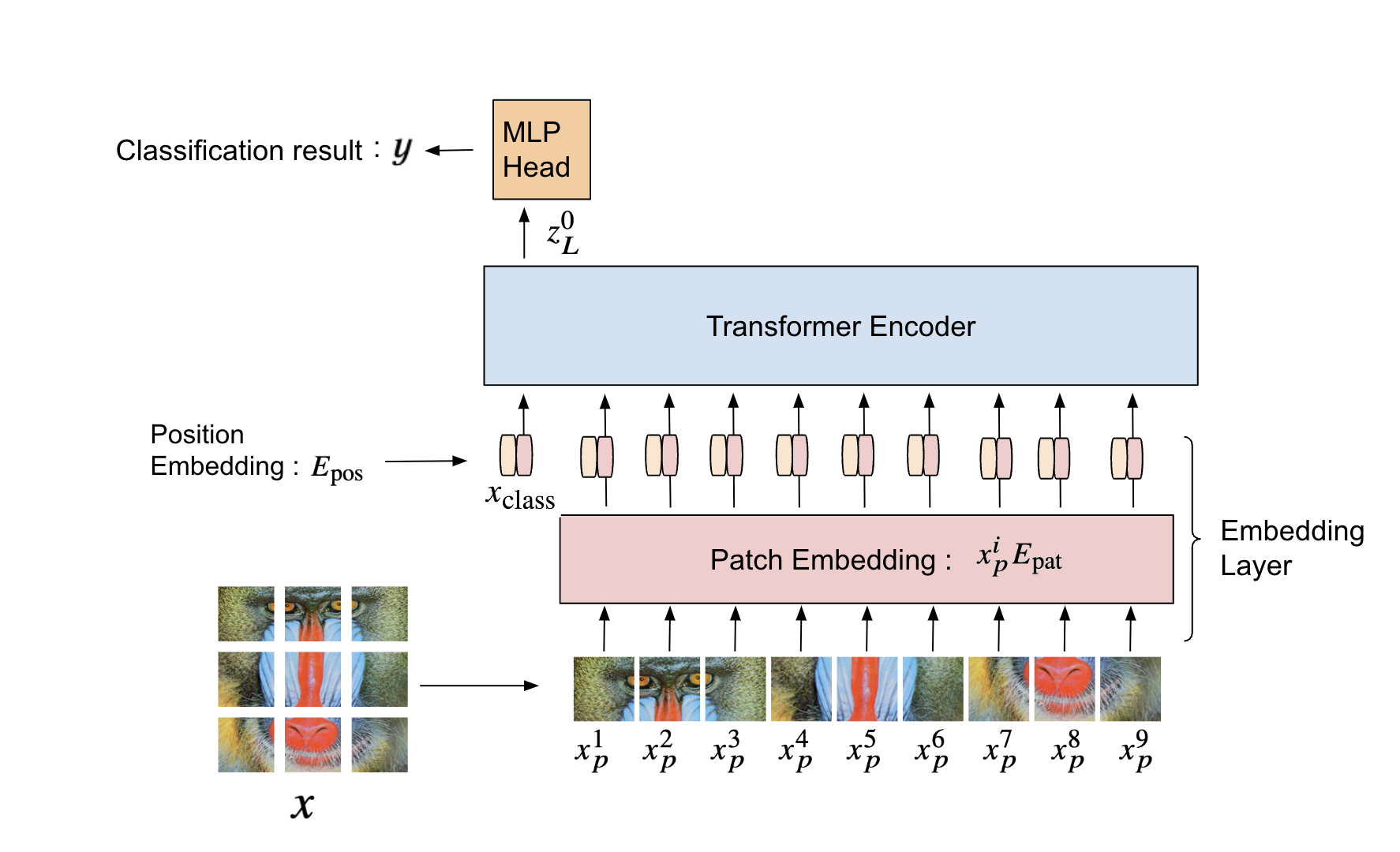}
    \caption{Overview of ViT}
\end{figure}
ViT is mainly used for image classification tasks and is known to have high classification performance\cite{ViT}. As shown in Figure 1, it consists of three components: embedding layer, transformer encoder, and MLP (Multi-Layer Perceptron) Head. In this paper, we focus on the embedding layer, which is a layer for converting an image into a feature vector.\\
\indent An input image $x\in{\mathbb{R}^{H \times W \times C}}$ is first divided into patches with a size of $P \times P$ where $H$, $W$, and $C$ are the height, width, and number of channels of the image. The number of patches $N$ is given as $N = (W/P) \times (H/P)$ as an integer. After that, each patch is flattened as $x_p^i = [x_p^i(1), x_p^i(2), ..., x_p^i(L)]$, where $L=P^2C$. Finally, a sequence of embedded patches is given as

\begin{align}
      Z_{0} &= [x_{\text{class}}; x_p^{1}E_{\text{pat}}; \cdots x_p^{i}E_{\text{pat}}; \cdots x_p^{N}E_{\text{pat}}] + E_{\text{pos}} 
\end{align}
where, 
\begin{align}
      E_{pos}= \left(  \left( e^{0}_{pos} \right)^T \cdots \left( e^{i}_{pos}  \right)^T  \cdots \left( e^{N}_{pos} \right)^T\right)^T, \nonumber \\  
      x_{class} \in{\mathbb{R}^{D}}, x_{p}^{i} \in{\mathbb{R}^{L}}, e_{pps}^{i} \in{\mathbb{R}^{D}},~~~~~~~~ \nonumber \\ 
      E \in{\mathbb{R}^{L \times D}}, E_{pos} \in{\mathbb{R}^{(N+1) \times D}}. ~~~~~~~~~~ \nonumber
\end{align}
$x_{class}$ is the classification token, $E_{pat}$ is the embedding (patch embedding) to linearly map each patch to dimensions $D$, $E_{pos}$ is the embedding (position embedding) that gives position information to patches in the image, $e^{0}_{pos}$ is the information of the classification token, and $e^{i}_{pos}$, $i=1,...,N$, is the position information of each patch.

\section{Proposed Method}
\subsection{Overview}

\begin{figure}[bth]
    \centering
    \includegraphics[bb=0 0 846 500,scale=0.35]{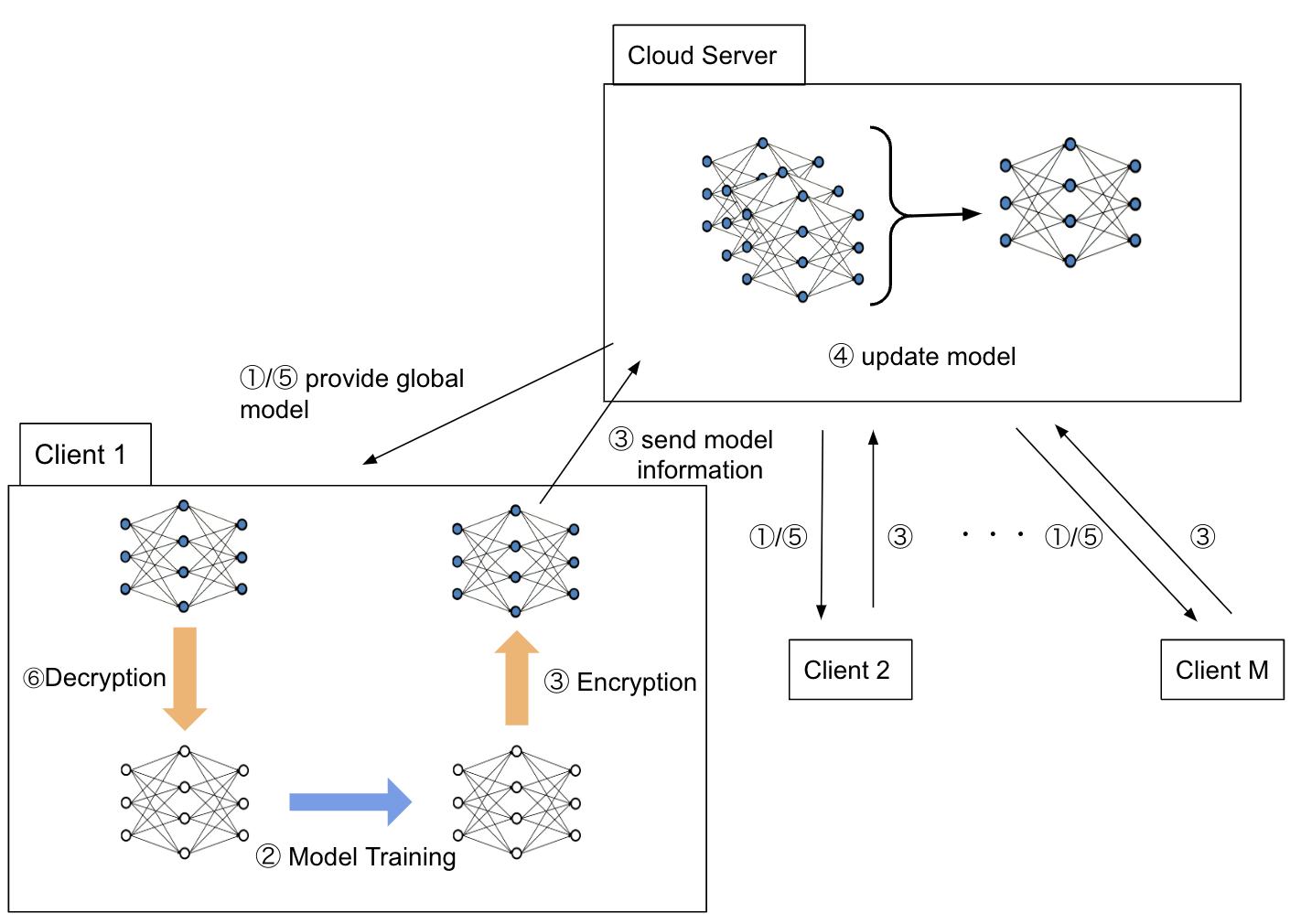}
    \caption{Framework of proposed method}
\end{figure}
The proposed method aims to prevent the visual information of plain training images from being restored from model parameters sent from each client to the server. Figure 2 shows the framework of the method, where we assume that ViT is used and a secret key for encryption is shared with all clients. The procedure of the method is summarized below.
\begin{quote}
 \begin{itemize}
  \item[\textcircled{\scriptsize{1}}] A server provider distributes an initial global model to all clients.
  \item[\textcircled{\scriptsize{2}}] Each client updates the global model with their training data.
  \item[\textcircled{\scriptsize{3}}] Each client encrypts the parameters of the updated model with a common key and sends it to the server.
  \item[\textcircled{\scriptsize{4}}] The server provider integrates the model information received from clients and updates the global model in the encrypted domain.
  \item[\textcircled{\scriptsize{5}}] The server provider sends the updated global model to all clients.
  \item[\textcircled{\scriptsize{6}}] Each client decrypts the global model with a common key. 
  \item[\textcircled{\scriptsize{7}}] Repeat \textcircled{\scriptsize{2}} to \textcircled{\scriptsize{6}} multiple times.
 \end{itemize}
\end{quote}
\indent In this framework, a malicious external third party and cloud provider (untrusted) have no secret key, so they cannot restore training data from updated model information sent from clients. The main contribution of this paper is to propose a method that allows us to update a global model in the encrypted domain for the first time.  \\

\subsection{Model Encryption}
Model encryption is carried out in \textcircled{\scriptsize{3}}.  In the method, patch embedding $E_{\text{pat}}$ and position embedding $E_{\text{pos}}$ in Eq.(1) are encrypted by using random matrices, respectively.
\subsubsection{Patch Embedding Encryption}
The following transformation matrix $E_{\text{a}}$ is used to encrypt patch embedding $E_{\text{pat}}$.
\begin{equation}
    E_\text{a} = 
    \begin{pmatrix}
        k_{(1,1)} & k_{(1,2)} & \cdots & k_{(1,L)} \\
        k_{(2,1)} & k_{(2,2)} & \cdots & k_{(2,L)} \\
        \vdots & \vdots & k_{(i,j)} & \vdots \\
        k_{(L,1)} & k_{(L,2)} & \cdots & k_{(L,L)}
    \end{pmatrix},
\end{equation}
where
\begin{align}
    E_\text{a} \in{\mathbb{R}^{L \times L}}, \text{det}E_\text{a} \neq 0, \nonumber \\
    k(i,j) \in{\mathbb{R}}, i,j \in \{1, ..., L\} \nonumber.
\end{align}
Note that the element values of $E_{\text{a}}$ are randomly decided, but $E_{\text{a}}$ has to have an inverse matrix.\\
Then, by multiplying $E_{\text{pat}}$ by $E_{\text{a}}$, an encrypted patch embedding $\widehat{E}_{\text{pat}}$ is given as
\begin{align}
    \widehat{E}_{\text{pat}} = E_{\text{a}} E_{\text{pat}}.
\end{align}
\subsubsection{Position Embedding Encryption}
Position Embedding $E_{\text{pos}}$ is encrypted as below.
\begin{quote}
 \begin{itemize}
  \item[1] Generate a random integer vector with a length of N as 
  \begin{align}
    l_t = [l_e(1), l_e(2), ..., l_e(i),..., l_e(N)],
  \end{align}
  where
  \begin{align}
     l_e(i) \in\{1,2,...,N \}, \nonumber \\
     l_e(i) \neq l_e(j) ~~~if ~~i \neq j, \nonumber \\
     i, j \in \{ 1,2,...,N\} \nonumber.
  \end{align}
  \item[2] Calculate $m_{(i,j)}$ as
  \begin{align}
    m_{(i,j)} =
    \begin{cases}
        1 & (j = l_e(i)) \\
        0 & (j \neq l_e(i)) .
    \end{cases}
  \end{align}
  \item[3] Define a random matrix as
\begin{align}
    E_\text{b} = 
    \begin{pmatrix}
        1 & 0 & 0 & \cdots & 0 \\
        0 & m_{(1,1)} & m_{(1,2)} & \cdots & m_{(1,N)} \\
        0 & m_{(2,1)} & m_{(2,2)} & \cdots & m_{(2,N)} \\
        \vdots & \vdots & \vdots & \ddots & \vdots \\
        0 & m_{(N,1)} & m_{(N,2)} & \cdots & m_{(N,N)} \\
    \end{pmatrix}
\end{align}
where
\begin{align}
    E_\text{b} \in{\mathbb{R}^{(N+1) \times (N+1)}} \nonumber.
\end{align}
.
  \item[4] Transform $E_{pos}$ to $\widehat{E}_{pos}$ as
\begin{align}
    \widehat{E}_{\text{pos}} = E_{\text{b}} E_{\text{pos}}.
\end{align}

 \end{itemize}
\end{quote}

\subsection{Global Model Update}
The cloud server updates the global model by using the model information received from each client. \\
For example, when using FedSGD, a global model is updated below.\\
\indent Let $M$ be the number of clients, $W^{(t)}$ be the parameters of the global model after $t$ updates, $\theta_{i}^{(t)}$ be the model update information (gradients) computed by client $i$ and $\tau$ be the learning rate. In this case, the global model is updated as follows;
\begin{gather}
  W^{(t+1)} = W^{(t)} - \tau \frac{1}{M} \sum^{M}_{i=1} \theta_{i}^{(t)}.
\end{gather}
Since the model is updated independently in each layer, model parameters in patch embedding and position embedding can be updated  from  Eq.(8) as
\begin{gather}
  W_{\text{pat}}^{(t+1)} = W_{\text{pat}}^{(t)} - \tau \frac{1}{M} \sum^{M}_{i=1} E_{\text{pat}, i}^{(t)} ~~ ,
\end{gather}
\begin{gather}
  W_{\text{pos}}^{(t+1)} = W_{\text{pos}}^{(t)} - \tau \frac{1}{M} \sum^{M}_{i=1} E_{\text{pos}, i}^{(t)}~.
\end{gather}
$E_{\text{pat},i}^{(t)}$ and $E_{\text{pos},i}^{(t)}$ are the parameters of patch and position embeddings updated by client $i$.\\
\indent According to Eqs.(3) and (9), the parameters of patch embedding are updated as
\begin{align}
   \widehat{W}_{\text{pat}}^{(t+1)} &= \widehat{W}_{\text{pat}}^{(t)} - \tau \frac{1}{M} \sum^{M}_{i=1} \widehat{E}_{\text{pat}, i}^{(t)} \nonumber \\
    &= E_{\text{a}} \left( W_{\text{pat}}^{(t)} - \tau \frac{1}{M} \sum^{M}_{i=1} E_{\text{pat}, i}^{(t)} \right) \nonumber \\
   &= E_{\text{a}} W_{\text{pat}}^{(t+1)} .
\end{align}
\indent According to Eqs.(7) and (10), the parameters of position embedding are updated as
\begin{align}
   \widehat{W}_{\text{pos}}^{(t+1)} &= \widehat{W}_{\text{pos}}^{(t)} - \tau \frac{1}{M} \sum^{M}_{i=1}  \widehat{E}_{\text{pos}, i}^{(t)} \nonumber \\
   &= E_{\text{b}} \left( W_{\text{pos}}^{(t)} - \tau \frac{1}{M} \sum^{M}_{i=1} E_{\text{pos}, i}^{(t)} \right) \nonumber \\
   &= E_{\text{b}} W_{\text{pos}}^{(t+1)} .
\end{align}
Eqs.(9) and (10) show that the global model on the cloud server can be updated in the encrypted domain.

\subsection{Model Decryption}
An encrypted global model is decrypted by each client as shown in \textcircled{\scriptsize{6}}. \\
\indent To decrypt patch embedding, $\widehat{W}_{\text{pat}}^{(t+1)}$ is multiplied by the inverse of matrix $E_{\text{a}}$ used for model encryption as
\begin{gather}
  E_{\text{a}}^{-1} \widehat{W}_{\text{pat}}^{(t+1)} = E_{\text{a}}^{-1} E_{\text{a}} W_{\text{pat}}^{(t+1)} = W_{\text{pat}}^{(t+1)}.
\end{gather}
In contrast, to decrypt position embedding, $\widehat{W}_{\text{pos}}^{(t+1)}$ is multiplied by the inverse of matrix $E_{\text{b}}$ used for model encryption as
\begin{gather}
  E_{\text{b}}^{-1} \widehat{W}_{\text{pos}}^{(t+1)} = E_{\text{b}}^{-1} E_{\text{b}} W_{\text{pos}}^{(t+1)} = W_{\text{pos}}^{(t+1)}
\end{gather}
From the above equations, we see that the updated parameters of the unencrypted model can be obtained from the updated parameters of the encrypted model. Accordingly, the proposed method is verified to obtain the same parameters as those of models trained without any encryption.

\section{Experiment Results}
A virtual server and five clients were set up on a single machine to verify the effectiveness of the proposed FL. All experiments were carried on an open source framework called Flower \cite{beutel2022flower}, and FedSGD\cite{pmlr-v54-mcmahan17a} was used as an aggregation algorithm. The CIFAR10 dataset, which consists of 50,000 training and 10,000 test color images with a seize of 32 × 32, was also used to fine-tune the ViT model pre-trained with Image-Net. In experiments, each client was given randomly selected 10,000 images as training data without duplicates, where images were resized from 32 × 32 × 3 to 224 × 224 × 3 to fit the size of images to that of ViT. We evaluated the classification accuracy by inputting 10,000 test images to the final global model. In the setting of ViT, the patch size P in patch embedding was set to 16, the number of split patches in an input image was N = 196, and the dimensionality of output feature vectors was D = 384.

\subsection{Classification Performance}

In an image classification task, we verified the model performance of the proposed method in terms of image classification accuracy, compared with a standard FL method without any encryption. \\
\indent Table 1 shows the comparison between the proposed method (encrypted) and the standard one (plain). From the table, the method was verified to maintain the same accuracy as the standard one. Accordingly, the proposed method did not cause any performance degradation even when using encryption. 

\begin{table}[h]
  \centering
  \caption{Classification Accuracy ($P=16, N=196, M=5$)}
\begin{tabular}{c|cc}
   & w/o Encryption & w/ Encryption (proposed) \\
  \hline
   accuracy (\%) & 89.77 & 89.77 \\
\end{tabular}
\end{table}
\subsection{Evaluation of Robustness against Restoration Attack}
To verify the effectiveness of the proposed method in terms of enhancing security, a visual information restoration attack was performed on the model information sent from each client. Attention privacy leakage attack (APRIL) \cite{Lu_2022_CVPR}, which is a method proposed for ViT and is known to restore the original image with high accuracy, was used as an attack method in the experiment. It was pointed out that the parameters of FL learned under the use of ViT have privacy vulnerability, and a method for analytically restoring images from model information was proposed in \cite{Lu_2022_CVPR}. 
\vspace{-30pt}
\begin{figure}[ht]
\hspace{-15pt}
\scalebox{0.75}[0.75]{
    \begin{tabular*}{50mm}{@{\extracolsep{\fill}}ccc}
        \begin{minipage}{4truecm}
             \centering
              \includegraphics[bb=-30 -30 389 389,scale=0.33]{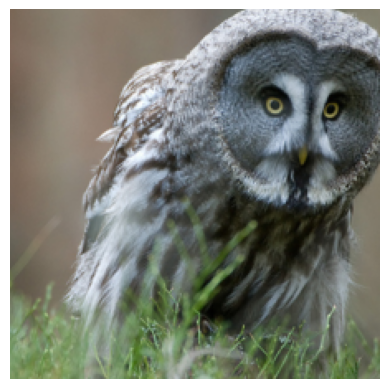}\\
            \end{minipage}
        &
        \begin{minipage}{4truecm}
             \centering
              \includegraphics[bb=-30 -30 389 389,scale=0.33]{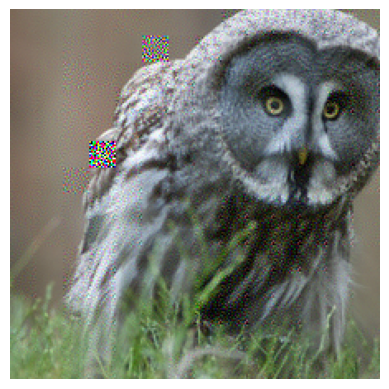}\\
            \end{minipage}
        &
        \begin{minipage}{4truecm}
             \centering
              \includegraphics[bb=-40 -40 389 389,scale=0.33]{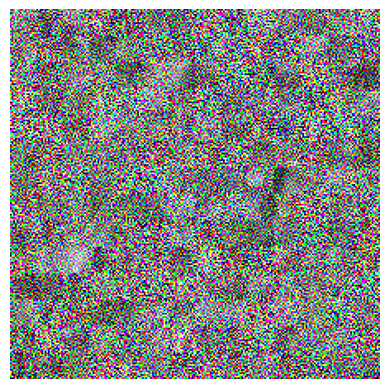}\\
            \end{minipage}\\
        \Large{(a)original image} & \Large{(b)without encrypted} & \Large{(c)proposed}  \\
    \end{tabular*}
    }
    \caption{Result of images restored from model parameter with APRIL}
\end{figure}
\\
\indent Figure 3 shows the results of the experiment. (a) is the original image, and (b) is the image reconstructed from model information (gradients) by APRIL attack in FL without any encryption. From the result, the visual information was confirmed to be restored. In contrast, (c) is the image restored from the model information protected by the proposed method. When applying the proposed method, the visual information was not restored by using APRIL.

\section{Conclusion}
In this paper, we proposed a novel framework of FL based on the embedded structure of ViT for enhancing the security of FL. In the proposed method, the model information shared between the cloud server and each client is encrypted with a secret key that the cloud provider does not know, and a global model is updated in the encrypted domain. In the experiments, the effectiveness of the proposed was confirmed in terms of image classification accuracy and robustness against an image restoration attack called APRIL.

\section*{Ackowledgment}
This study was partially supported by JSPS KAKENHI (Grant Number JP21H01327) and JST CREST (Grant Number JPMJCR20D3).
\bibliographystyle{IEEEtrans}
\bibliography{main}

\vspace{12pt}

\end{document}